\documentclass[twocolumn,amsmath,notitlepage,aps,prl,10pt,superscriptaddress,floatfix,letterpaper,reprint
]{revtex4-2}
\usepackage{hyperref}
\usepackage[usenames]{color}
\usepackage{amssymb}
\usepackage{graphicx}
\usepackage{physics}
\usepackage{commath}

\begin{document}

\title{Nevanlinna Analytical Continuation}

\author{Jiani Fei}
\affiliation{%
 Department of Physics, University of Michigan, Ann Arbor, Michigan 48109, USA
}%
\author{Chia-Nan Yeh}
\affiliation{%
 Department of Physics, University of Michigan, Ann Arbor, Michigan 48109, USA
}%
\author{Emanuel Gull}%
\affiliation{%
 Department of Physics, University of Michigan, Ann Arbor, Michigan 48109, USA
}%

\date{\today}

\begin{abstract}
Simulations of finite temperature quantum systems provide imaginary frequency Green's functions that correspond one-to-one to experimentally measurable real-frequency spectral functions.
However, due to the bad conditioning of the continuation transform from imaginary  to real frequencies, established methods tend to either wash out spectral features at high frequencies or produce spectral functions with unphysical negative parts.
Here, we show that explicitly respecting the analytic `Nevanlinna' structure of the Green's function leads to intrinsically positive and normalized spectral functions, and we present a continued fraction expansion that yields all possible functions consistent with the analytic structure.
Application to synthetic trial data shows that sharp, smooth, and multi-peak data is resolved accurately. Application to the band structure of silicon demonstrates that high energy features are resolved precisely. Continuations in a realistic correlated setup reveal additional features that were previously unresolved.
By substantially increasing the resolution of real frequency calculations our work overcomes one of the main limitations of finite-temperature quantum simulations.
\end{abstract}

\maketitle

The central object of finite-temperature field theories is the Matsubara Green's function $\mathcal{G}(i\omega_n)$. This quantity corresponds to the retarded Green's function $G^R(\omega)$ and the spectral function $A(\omega)=-\frac{1}{\pi} \text{Im} G^R(\omega)$, which characterizes the single-particle excitation spectrum measurable by photoemission spectroscopy.
Finite temperature simulations ranging from perturbative calculations \cite{Hedin65,Dahlen05,Phillips14} to lattice \cite{Blankenbecler81} and continuous-time \cite{Gull11} quantum Monte Carlo and lattice QCD  \cite{Asakawa01,Tripolt19,Rothkopf20} simulations obtain the Matsubara Green's function and analytically continue it in post-processing to obtain spectral information.

As the kernel relating $\mathcal{G}(i\omega_n)$ to $G^R(\omega)$ is ill conditioned \cite{Jarrell96}, a direct inversion of the relation is infeasible in practice. Instead, continuation methods such as a Pad\'{e} continued fraction fit \cite {Baker96} of Matsubara data \cite{Vidberg77,Beach00,Ostlin12,Osolin13,Schott16,Han17}, the Maximum Entropy (MaxEnt) method \cite{Bryan90,Creffield95,Jarrell96,Beach04,Gunnarsson10,Bergeron16,Levy17,Gaenko17,Kraberger17,Rumetshofer19,Sim18}, or stochastic analytic continuation (SAC) and variants \cite{Sandvik98,Mishchenko00,Gunnarsson07,Fuchs10,Goulko17,Otsuki17,Krivenko19} are employed.

MaxEnt and SAC aim to fit, rather than interpolate, the spectral function to Matsubara data consistent with specified error bars. The methods are generally successful for noisy data but struggle to resolve high frequency information, sharp peaks, and spectral functions with multiple features. In contrast, Pad\'{e} methods provide a rational interpolation of the Matsubara data. While sharp features appear, Pad\'{e} spectral functions typically change sign and do not satisfy the proper normalization and moment structure, making them ill suited for analyzing information away from low frequency.

\begin{figure}[bth]
\begin{center}
\includegraphics[width=0.99\columnwidth]{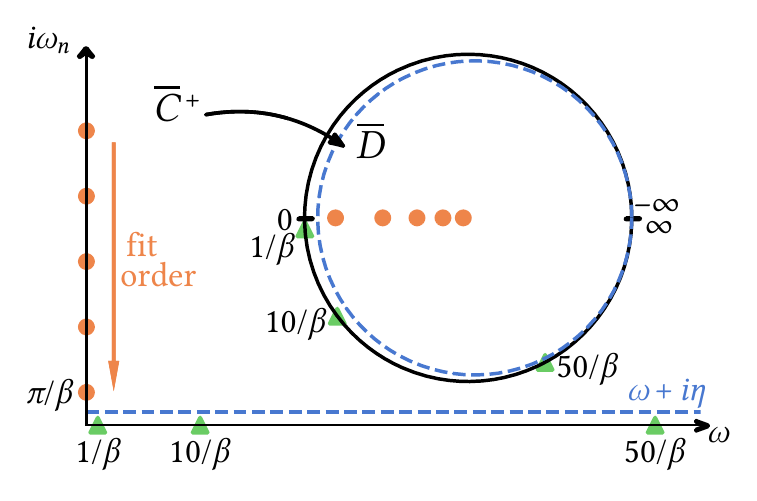}
\end{center}\vspace{-1cm}
\caption{Analytic continuation setup with fermion Matsubara points at  $i\omega_n$ and real frequency axis $\omega$. The retarded Green's function is evaluated $\eta$ (small) above the real axis. Inset: M\"{o}bius transform of the closed upper half plane $\overline{\mathcal{C}^+}$ to the closed unit disk $\overline{\mathcal{D}}$.}\label{fig:ComplexPlaneSketch}
\end{figure}

In this paper, we show that a continued fraction interpolation of Matsubara data can overcome all of these problems, provided that the interpolant satisfies the correct analytic structure. Spectral functions are then positive and respect analytically known moments, they are able to resolve both smooth broad features and sharp peaks, they can resolve sharp features at very high energy, and they can resolve multiple spectral features accurately. While there is an infinite number of valid interpolants, this freedom can be precisely characterized and used to find the `best possible' spectral function by optimizing a functional norm. 

Our continued fractions pave the way to analytic continuations with substantially better frequency resolution, devoid of the deficiencies of previous attempts, especially in situations where high precision Matsubara data is available.
The remainder of this paper will introduce the analytic structure of Green's functions, derive an interpolation algorithm that respects this analytic structure, and characterize the degrees of freedom in the interpolation. We will then present results for synthetic benchmark Green's functions (with emphasis on trial functions exhibiting both sharp and smooth features), the band structure of silicon (with emphasis on sharp high frequency features), and correlated real-materials simulations.

The retarded Green's function $G^R$ is analytic in the upper half of the complex plane, $\mathcal{C}^+$,  and contains singularities in the lower half plane. The Matsubara Green's function $\mathcal{G}(i\omega_n)$ and the retarded Green's function $G^R(\omega+i\eta)$ can be expressed consistently by replacing the variables $i\omega$ and $\omega+ i\eta$ with a single complex variable $z$. Analytic continuation is used to obtain $G^R$ from $\mathcal{G}$.

In complex analysis, a Nevanlinna function 
 is a complex function which is analytic in the open upper half plane $\mathcal{C}^+$ and has non-negative imaginary part, i.e. maps into $\overline{\mathcal{C}^+}$ (the overline denotes inclusion of the boundary). Denoting the class of Nevanlinna functions as $\mathtt{N}$, the negative of the Green's function  $\mathcal{NG} = -\mathcal{G}$ restricted to $\mathcal{C}^+$ is a Nevanlinna function, {\it i.e.} $\mathcal{NG}: \mathcal{C}^+ \to \overline{\mathcal{C}^+}$ and $\mathcal{NG}\in \mathtt{N}$.

This follows directly from the Lehmann representation of the Green's function $\mathcal{G}$,
\begin{align}
	\mathcal{G}(\gamma, z) &= \frac{1}{Z}\sum_{m, n} \frac{\lvert\langle m \lvert c_\gamma ^{\dagger}\rvert n \rangle\rvert^2}{z + \textit{E}_n - \textit{E}_m}(e^{-\beta \textit{E}_n}+e^{-\beta \textit{E}_m})\label{eqn: Lehmann} 
\end{align}
where $\textit{E}_m$ and $\textit{E}_n$ are eigenvalues corresponding to the eigenstates $\lvert m \rangle$ and $\lvert n \rangle$ of a Hamiltonian, $Z$ is the partition function, $\beta$ is the inverse temperature, and $c_\gamma ^{\dagger}$ is the creation operator for orbital $\gamma$. Defining
$Q = \frac{1}{Z} \lvert\langle m \lvert c_\gamma ^{\dagger}\rvert n \rangle\rvert^2 (e^{-\beta \textit{E}_n}+e^{-\beta \textit{E}_m}) \geqslant 0$
and setting $z = x + yi$ with $y > 0$, {\it i.e.} $z\in \mathcal{C}^+$,
\begin{align}
	\mathcal{G}(\gamma, z) 
	= \sum_{m,n}\frac{Q(x + \textit{E}_n - \textit{E}_m-yi)}{(x+ \textit{E}_n - \textit{E}_m)^2+y^2}
\end{align}
and thus
\begin{align}
	\Im{\mathcal{G}(\gamma, z)} &= -\sum_{m,n}\frac{Qy}{(x+ \textit{E}_n - \textit{E}_m)^2+y^2} \leqslant 0,
\end{align}
implying that $\mathcal{NG} \in \mathtt{N}$.

To perform analytic continuation from the Matsubara to the real axis, we aim to find an interpolant for $\mathcal{NG}$ in the class of Nevanlinna functions $\mathtt{N}$, rather than a generic continued fraction. By construction, this function will pass through all Matsubara points (see Fig.~\ref{fig:ComplexPlaneSketch}) and have a positive imaginary part in the upper half plane, including just above the real axis. Spectral functions $A(\omega) = \lim_{\eta\to 0^+}\frac{1}{\pi}\text{Im}\{\mathcal{NG}(\omega+i\eta)\} $ are therefore intrinsically positive, avoiding the common failure of Pad\'{e} interpolants.



We construct Nevanlinna interpolants using the Schur algorithm \cite{Schur}, which is originally a continued fraction expansion for all holomorphic disk functions mapping from $\mathcal{D}$ to $\overline{\mathcal{D}}$, where $\mathcal{D} = \{z:\lvert z\rvert < 1\}$ is the open unit disk in the complex plane, $\overline{D}$ the closed unit disk. Schur algorithm is modified to expand all contractive functions \cite{Adamyan2003}, which are holomorphic functions mapping from $\mathcal{C}^+$ to $\overline{\mathcal{D}}$. The invertible M\"{o}bius transform $h: \overline{\mathcal{C}^+} \to \overline{\mathcal{D}}, z\mapsto\frac{z-i}{z+i}$ on function value (with half-plane domain unchanged) maps Nevanlinna functions one-to-one to contractive functions (see Fig.~\ref{fig:ComplexPlaneSketch}). The Nevanlinna interpolation problem is therefore mapped into the problem of constructing the contractive function $\theta$ which is M\"{o}bius transformed from $\mathcal{NG}$,
\begin{align}
	\theta(Y_i) = \lambda_i = h(C_i) = \frac{C_i-i}{C_i+i}\quad i=1,2,\dots,M\label{contractiveProblem}
\end{align}
where $Y_i$ is the i-th Matsubara frequency, $C_i$ is the value of $\mathcal{NG}$ at $Y_i$, and $\lambda_i$ is the value of $\theta$ at $Y_i$.

Remarkably, there is a straightforwardly verifiable criterion for the existence of Nevanlinna interpolants directly based on input data, which is a generalization of the Pick criterion \cite{Pick,Tannenbaum17}. Nevanlinna interpolants exist if and only if the Pick matrix,
\begin{align}
	\left[\frac{1-\lambda_i\lambda_j^*}{1-h(Y_i)h(Y_j)^*}\right]_{i,j}\quad i,j=1,2,\dots,M\label{existCondition}
\end{align}
is positive semi-definite; and a unique solution only if it is singular. In practice, we find that most noisy data (in particular most Monte Carlo data) does not satisfy this criterion, meaning that there does not exist a globally positive and holomorphic function in the upper half plane that passes through all Matsubara points.

The iterative construction of contractive interpolant comes as follows. First, note that a contractive function $\theta(z)$,
\begin{align}
	\theta(z)&=\frac{\frac{z-Y_1}{z-Y_1^*}\tilde\theta(z)+\gamma_1}{\gamma_1^*\frac{z-Y_1}{z-Y_1^*}\tilde\theta(z)+1}\label{eqn:inter1}
\end{align}
will satisfy $\theta(Y_1)=\gamma_1$ for any contractive function $\tilde\theta(z)$ \cite{Adamyan2003}. Given an interpolation problem for $j$ nodes, $\theta(Y_k)=\gamma_k, k=1\dots j$, Eq.~\ref{eqn:inter1} defines an interpolation problem for the $j-1$ nodes $Y_2, \dots, Y_{j}$ for $\tilde \theta$. The equation results from Schur's expansion for any disk function with known value $\gamma_1$ at the origin \cite{Ammar87} by the conformal map $g: \mathcal{C}^+ \to \mathcal{D}, z\mapsto\frac{z-Y_j}{z-Y_j^*}$, projecting the origin in $\mathcal{D}$ to $Y_1$ in $\mathcal{C}^+$.

Eq.~\ref{eqn:inter1} suggests an iterative algorithm starting from the original interpolation problem for $\theta_1=\theta$ of $M$ points, which defines an interpolation problem $\theta_2$ for $M-1$ points, which in turn defines an interpolation problem $\theta_3$ for $M-2$ points, etc. Concatenating these interpolation problems results in a continued fraction form for $\theta$. Denoting $\theta_j(Y_j)=\phi_j$ to be the point ignored by the interpolant $\theta_{j+1}$, there is a freedom to choose an arbitrary contractive function $\theta_{M+1}$ at the last step that is reduced from $\theta_M(Y_M)=\phi_M$ fulfilling the interpolation problem. As we will show below, this freedom can be used to satisfy additional criteria, such as smoothness.

The recursive final $\theta$ can conveniently be written in a matrix form \cite{Adamyan2003,Ammar87},
\begin{align}
	\theta(z)[z;\theta_{M+1}(z)]=\frac{a(z)\theta_{M+1}(z)+b(z)}{c(z)\theta_{M+1}(z)+d(z)}\label{eqn:SchurFinal}
\end{align}
where
\begin{align}
	\begin{pmatrix}
		a(z) & b(z)\\
		c(z) & d(z)
	\end{pmatrix}
	=\prod_{j=1}^{M}
	\begin{pmatrix}
		\frac{z-Y_j}{z-Y_J^*} & \phi_j\\
		\phi_j^*\frac{z-Y_j}{z-Y_j^*} & 1
	\end{pmatrix}.	
\end{align}
with $j$ increasing from left to right. $\theta$ is then back transformed to a Nevanlinna interpolant via the inverse M\"{o}bius transform $h^{-1}$,
$\mathcal{NG}(z)=h^{-1}(\theta(z))=i\frac{1+\theta(z)}{1-\theta(z)}$. In practice, we found that solving these equations required at least quadruple precision.

\begin{figure}[tb]
\begin{center}
\includegraphics[width=0.97\columnwidth]{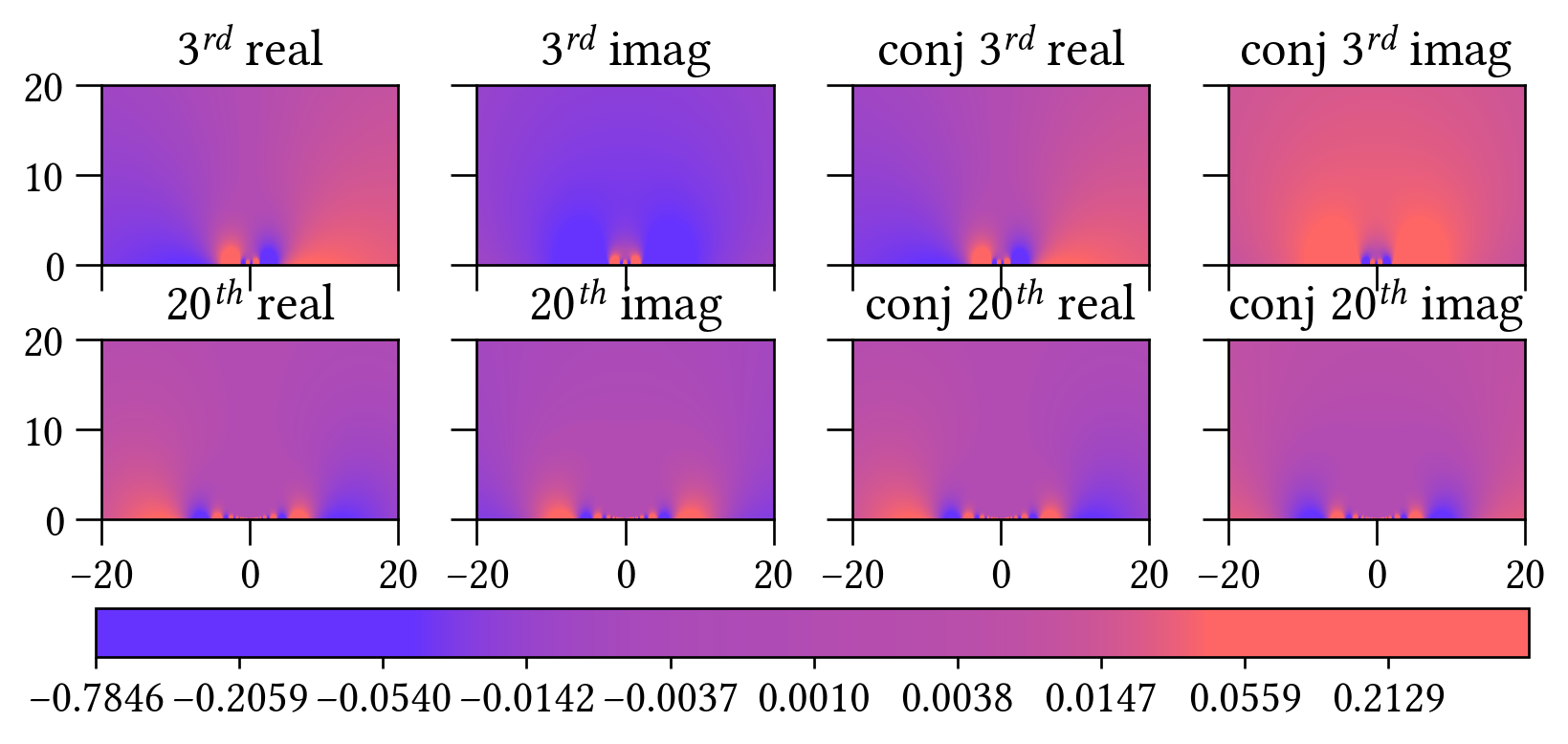}
\end{center} \vspace{-0.5cm}
\caption{Real and imaginary parts of the $3$rd and $20$th Hardy functions and conjugates used in the optimization, plotted in the upper half complex plane.}\label{fig:hardyBasis}
\end{figure}

The choice of $\theta_{M+1}$ is still to be discussed. Any contractive function will yield a valid interpolation and spectral function, and therefore this freedom can be used to select the `best' of all consistent spectral functions. It is natural to expand $\theta_{M+1}$ into a set of basis functions and optimize the resulting spectral function in some norm as a function of those basis function coefficients.
As we demonstrate below, a constant $\theta_{M+1}$ results in spectral functions with oscillations. We therefore employ the freedom in choosing $\theta_{M+1}$ to eliminate these oscillations and obtain the smoothest possible spectral function. Other criteria, such as proximity to a trial function that is either featureless or exhibits a desired feature, are possible but have not been pursued here.

The Hardy space $H^2$  \cite{Marvin} in $\mathcal{C}^+$ consists of holomorphic functions whose mean square value on $\omega+i\eta$ remains bounded as $\eta\downarrow 0$. The Hardy basis $f^k(z)=\frac{1}{\sqrt{\pi}(z+i)}\left(\frac{z-i}{z+i}\right)^k$ and its conjugate (see Fig.~\ref{fig:hardyBasis}) generate functions in the contractive function space with rapid variations at low frequency. Expanding $\theta_{M+1}=\sum_{k=0}^H a_k f^k(z)+ b_k (f^{k}(z))^*$   allows to determine the complex coefficients $a_k$ and $b_k$ by minimizing a smoothness norm such as $F[A_{\theta_{M+1}}(\omega)]=|1-\int A_{\theta_{M+1}}(\omega)|^2 +\lambda \int A_{\theta_{M+1}}''(\omega)^2$, where the first term enforces proper normalization while the second term promotes smoothness by minimizing second derivatives (we typically use $\lambda=10^{-4}$ and $H=25$). In our implementation, we used a conjugate gradient minimizer of the Dakota package \cite{Adams14} to minimize the norm and eliminate oscillations from the spectral function.

Finally, we remark that the moments of the Green's function, which are known analytically from commutator expansions \cite{Rusakov14}, can be enforced explicitly. This is known as the Hamburger moment problem \cite{Hamburger20}, and combinations with the Schur algorithm are straightforward \cite{Akhiezer}. In our simulations, which predominantly used data accurate to double precision, we found that non-uniform grids \cite{Li20,Kresse20} with data at large Matsubara frequencies contained enough moment information that an explicit enforcement of the moments did not yield any advantage in practice. A combination may become useful if fits to noisy Monte Carlo data are attempted.

\begin{figure}[tb]
\begin{center}
\includegraphics[width=0.99\columnwidth]{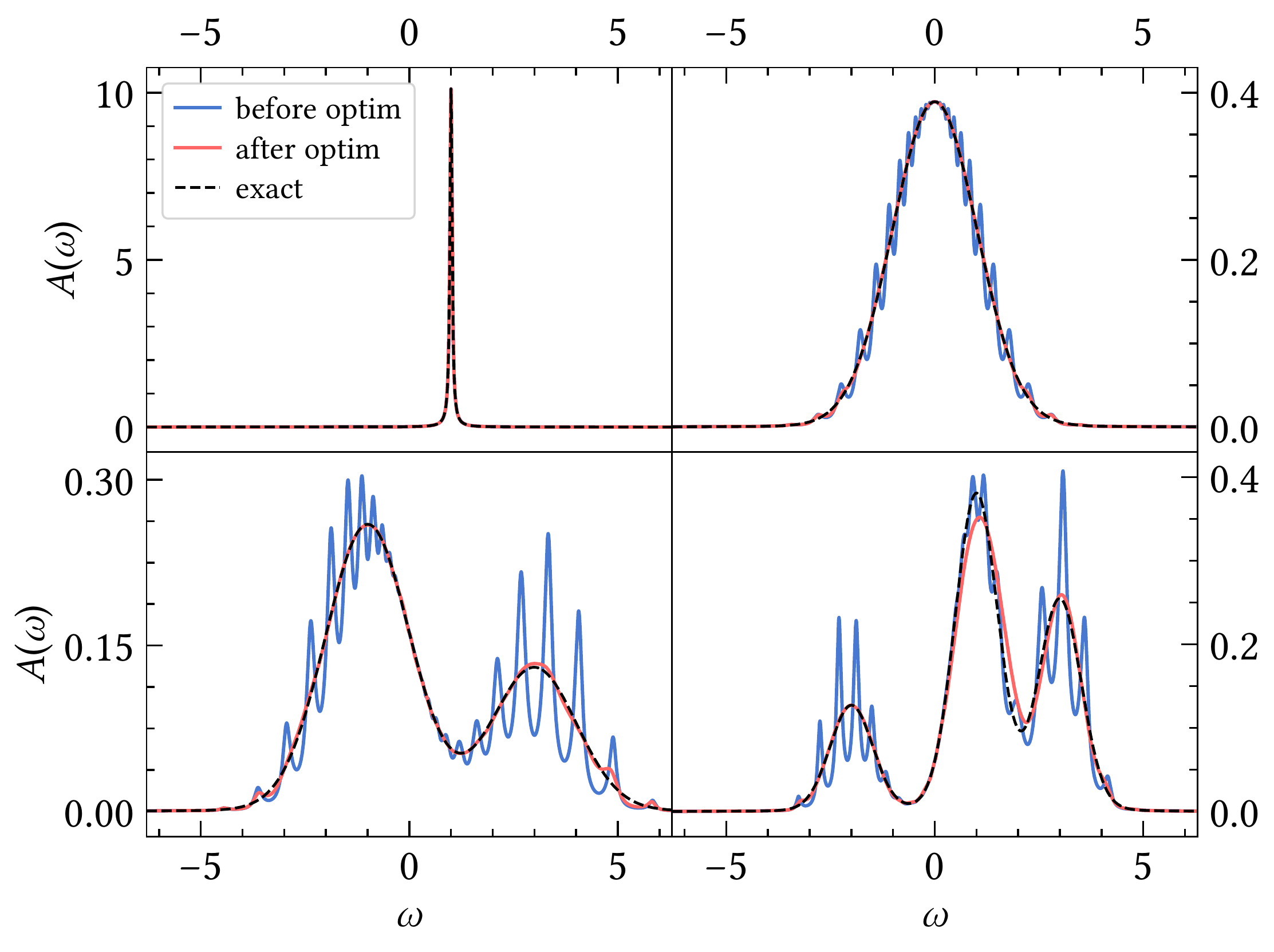}
\end{center}\vspace{-0.7cm}
\caption{Continuation with and without Hardy function optimization. Off-centered $\delta$ peak (top left), Gaussian (top right), two-peak scenario (bottom left), and a three-peak scenario (bottom right). $\beta=100$, IR grid \cite{Shinaoka17,Li20} with $36$ Matsubara positive frequency points.\label{fig:hardySmoothing}}
\end{figure}

Fig.~\ref{fig:hardySmoothing} shows the results of the method for four prototypical spectral functions: an off-center $\delta$-peak `level' (top left), a centered Gaussian (top right), a double-peak `pseudogap' scenario (bottom left), and a three-peak structure with a second, smaller peak hidden behind the first peak. Black lines show input data that is then back-continued to the imaginary axis in double precision as an input for the interpolation algorithm.

We show two sets of results from Nevanlinna continuation which both interpolate all Matsubara points and are intrinsically positive and normalized. First, the result of an interpolation using a constant function $\theta_{M+1}=0$. For the $\delta$-function, the interpolation is very close to the original data. However, other curves display artificial oscillations. The number of these oscillations increases as additional Matsubara points are fit. Nevertheless, the approximate shape of the original spectral function is evident in all interpolations.

Next, we exploit the additional freedom to find the `best' function among all possible interpolants by minimizing the functional $F[A_{\theta_{M+1}}(\omega)]$ with $25$ Hardy basis coefficients and their conjugates. Other choices of functionals, including minimizing $\int A^2(\omega)$ while keeping $\int A(\omega)$ constant, yield similar results. As is evident in Fig.~\ref{fig:hardySmoothing}, the minimization eliminates all oscillations and produces a spectral function that is both smooth and very close to the original data, while not destroying the sharp features of the $\delta$ peak in the top left panel.

\begin{figure}[tb]
\begin{center}
\includegraphics[width=0.95\columnwidth]{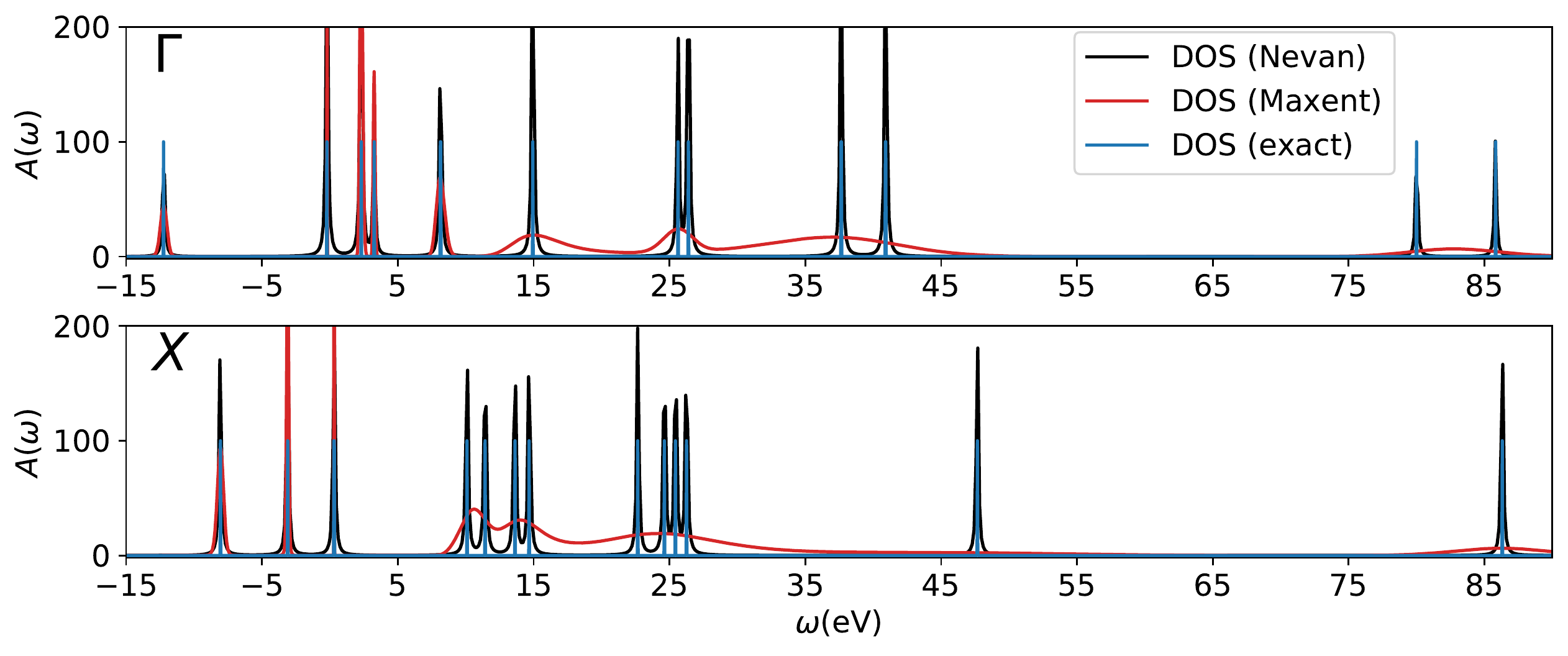}
\end{center}\vspace{-0.75cm}
\caption{LDA band structure (Kohn Sham eigenvalues, DOS) of solid Si (green) at the $\Gamma$ and the $X$ point, as well as Nevanlinna (blue) and MaxEnt (orange) continuations of the corresponding Green's functions. T=316 K, $52$ non-uniform  \cite{Li20} IR Basis \cite{Shinaoka17} Matsubara positive frequency points.}\label{fig:Si}
\end{figure}

We now turn to Fig.~\ref{fig:Si}. Shown are $k$-resolved DFT Kohn Sham eigenvalues (the `band structure') of solid Si in the LDA approximation at the $\Gamma$ and the $X$ point, obtained on an $8\times8\times8$ grid in the \emph{gth-dzvp-molopt-sr} basis \cite{VandeVondele07} with \emph{gth-pbe} pseudopotential \cite{Goedecker96}. The eigenvalue spectrum at the $k$-points shown is back-continued to the Matsubara axis in double precision, at  $T=316 K$ and with $52$ non-uniform  \cite{Li20}  IR Basis  \cite{Shinaoka17} Matsubara points, for each orbital individually, and then analytically continued. Shown are Nevanlinna  (blue) and Maximum Entropy continuations (orange). It is evident that Nevanlinna resolves the delta peaks at the right locations, even at very high energy, whereas MaxEnt only obtains the approximate area, but not the sharp unperturbed levels at high energy.
Continuations of this type often appear in correlated simulations of real materials, where the spectral function broadening due to electron correlations needs to be distinguished from a broadening due to analytic continuation deficiencies. Our method, which is able to capture both broad features near the Fermi energy and sharp features away from it,  therefore offers the unique capability of accurately resolving bandstructure at high energy. The fact that sharp features are resolved, despite Hardy function smoothing, hints at the severe restriction of the functions available within the Nevanlinna space.

\begin{figure}[tb]
\begin{center}
\includegraphics[width=0.95\columnwidth]{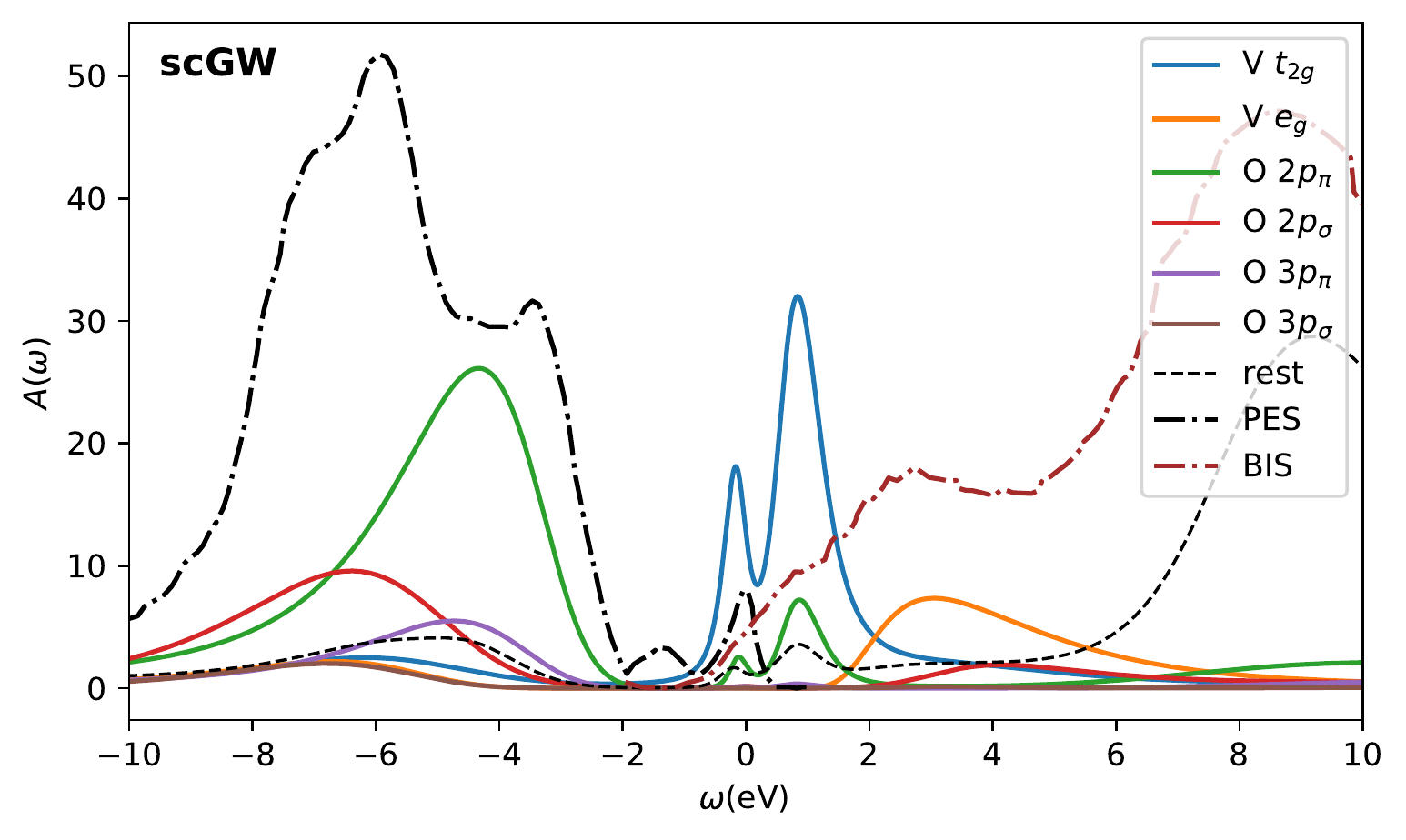}\\
\includegraphics[width=0.95\columnwidth]{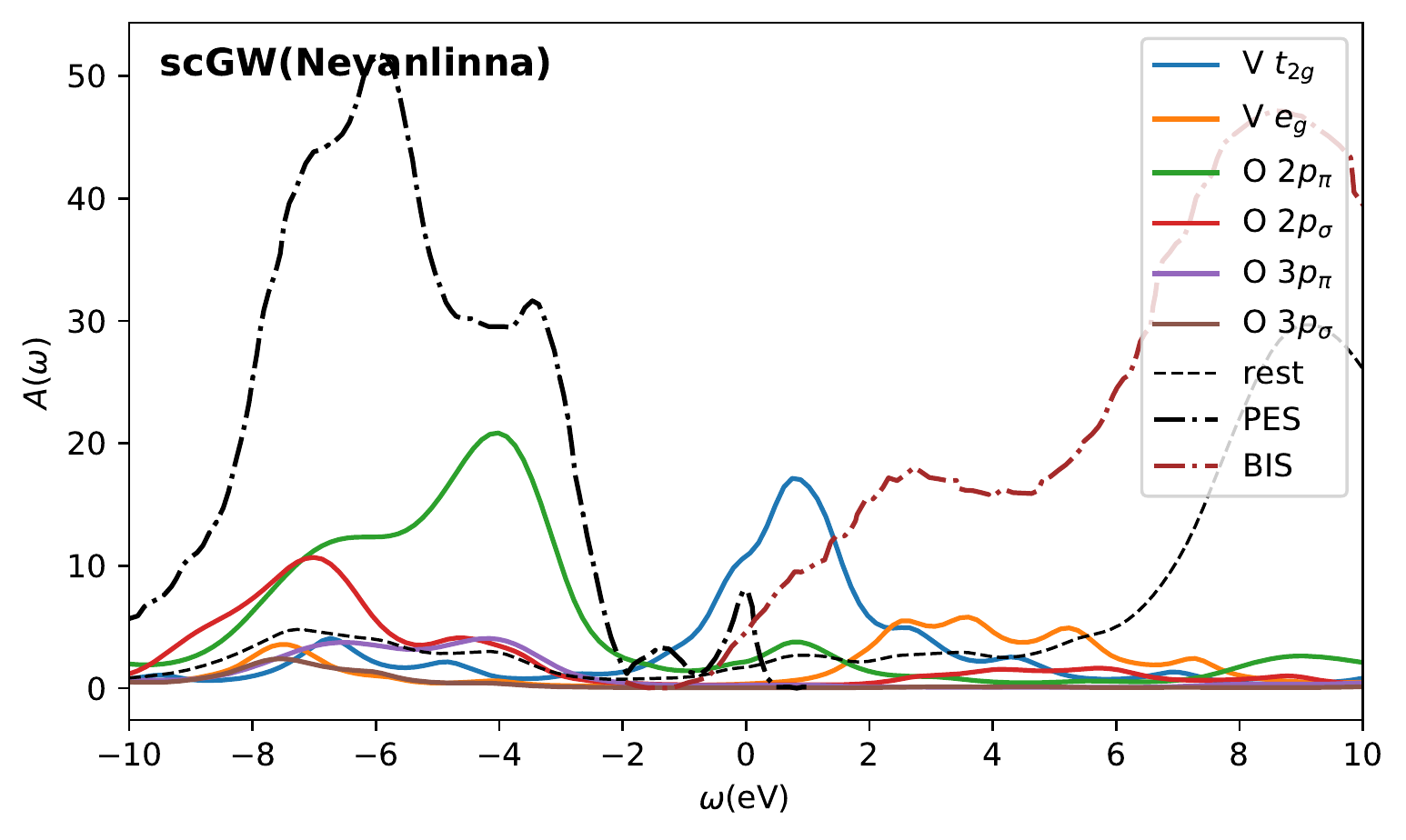}
\end{center}
\caption{Orbital-resolved realistic band structure of SrVO$_3$ from self-consistent GW continued with MaxEnt (top) and with Nevanlinna (bottom) \cite{Yeh20}.}\label{fig:SrVO3}
\end{figure}

To illustrate the power of the method in a difficult realistic correlated setting, we show near-Fermi-energy results from a self-consistent GW \cite{Hedin65} calculation of SrVO$_3$ in Fig.~\ref{fig:SrVO3}, from MaxEnt (top) \cite{Levy17} and Nevanlinna (bottom panel). For methods details and physics discussion see Ref.~\cite{Yeh20}. Shown are experimental photoemission \cite{Yoshimatsu10} and bremsstrahlung isochromat spectroscopy \cite{Morikawa95} data along with orbitally resolved local GW spectra obtained at $T=1579K$ on a $6\times6\times6$ grid in a Gaussian \emph{gth-dzvp-molopt-sr} basis  \cite{VandeVondele07} with \emph{gth-pbe} pseudopotential \cite{Goedecker96} at $84$ frequency points. The four-fermion Coulomb integrals are decomposed into a combination of auxiliary even-tempered Gaussian for Strontium and \emph{def2-svp-ri}~\cite{Hattig05} bases for all other atoms. Both methods recover the same overall features. However, Nevanlinna continuation reveals additional details such as multiplet structures in the occupied and unoccupied bands, and does not exhibit artificial oscillations in the t$_{2g}$ bands near the Fermi energy.

In conclusion, we have derived a continuous fraction expansion for analytic continuation of Matsubara data. The expansion constructs a class of functions that intrinsically respect the analytic structure of Green's functions. By construction, the functions are positive, consistent with the Matsubara data, and respect the moment information contained in the input data. We have provided a parametrization of the class of all possible spectral functions in terms of contractive functions, which can be expanded into Hardy basis. We have then shown how optimizations in this space of functions (e.g. to obtain the `smoothest' function consistent with the input data) are possible and yield the expected result.

An application to synthetic benchmark data showed that the method could resolve both sharp and smooth features. An application to the band structure of silicon showed that high energy features are precisely resolved. Finally, an application to correlated real materials simulation revealed additional structure that was not visible in a MaxEnt continuation.

Our description is accompanied by a sample implementation \footnote{See supplemental materials}. Natural extensions to be considered in the future are the calculation of self-energy continuations \cite{Wang09}, matrix valued functions \cite{Kraberger17}, anomalous functions \cite{Gull15} and most importantly bosonic response functions \cite{Jarrell96} such as optical conductivities \cite{Gunnarsson10}.

\begin{acknowledgments}
This work was supported by the Simons Foundation via the Simons Collaboration on the Many-Electron Problem and NSF DMR 2001465. We thank Sergei Iskakov for help with non-equidistant grids.
\end{acknowledgments}
\bibliographystyle{apsrev4-2}
\bibliography{refs}
\end{document}